\documentclass[iop]{emulateapj}

\usepackage{natbib,aas_macros,amsmath}
\usepackage{multirow,color}

\begin{document}
\slugcomment{Accepted for publication in The Astrophysical Journal}
\shorttitle{BCG progenitor candidate at $z\sim8$}
\shortauthors{Ishigaki et al.}

\title{
Very Compact Dense Galaxy Overdensity with $\delta \simeq 130$ Identified at $\lowercase{z}\sim 8$:\\
Implications for Early Protocluster and Cluster-Core Formation
}

\author{Masafumi Ishigaki\altaffilmark{1,2},  Masami Ouchi\altaffilmark{1,3}, and Yuichi Harikane\altaffilmark{1,2}}
\email{ishigaki@icrr.u-tokyo.ac.jp}
\altaffiltext{1}{Institute for Cosmic Ray Research, The University of Tokyo, Kashiwa, Chiba 277-8582, Japan}
\altaffiltext{2}{Department of Physics, University of Tokyo, 7-3-1 Hongo, Bunkyo-ku, Tokyo 113-0033, Japan}
\altaffiltext{3}{Kavli Institute for the Physics and Mathematics of the Universe (Kavli IPMU, WPI), University of Tokyo, Kashiwa, Chiba 277-8583, Japan}

\begin{abstract}
We report the first identification of a compact dense galaxy overdensity at $z\sim8$
dubbed A2744z8OD. A2744z8OD consists of eight $Y$-dropout galaxies behind Abell 2744
that is originally pinpointed by Hubble Frontier Fields studies.
However, no studies have, so far, derived basic physical quantities
of structure formation or made comparisons with theoretical models.
We obtain a homogeneous sample of dropout galaxies at $z\sim8$
from eight field data of Hubble legacy images that are as deep as the A2744z8OD data.
Using the sample, we find that a galaxy surface overdensity value of A2744z8OD is very high $\delta\simeq130$, where $\delta$ is
defined by an overdensity in a small circle of $6''$ ($\simeq30$ physical kpc) radius.
Because there is no such a large $\delta$ value reported for high-$z$ overdensities to date,
A2744z8OD is a system clearly different from those found in previous high-$z$ overdensity studies.
In the galaxy+structure formation models of Henriques et al. (2015),
there exist a very similar overdensity, Modelz8OD,
that is made of eight model dropout galaxies at $z\sim8$ in a $6''$-radius circle.
Modelz8OD is a progenitor of a today's $10^{14}M_\odot$ cluster, and
more than a half of the seven Modelz8OD galaxies are merged into the brightest cluster galaxy (BCG) 
of the cluster. 
If Modelz8OD is a counterpart of A2744z8OD, the models suggest that
A2744z8OD would be a part of a forming cluster core of a today's $10^{14}M_\odot$ cluster
that started star formation at $z>12$.
\end{abstract}


\section{Introduction} \label{sec:Introduction}

Hubble Space Telescope (HST) observations identify faint high-redshift galaxies at the epoch of reionization and beyond.
Hubble Ultra Deep Field 2012 (UDF12) project has revealed the faint end of the UV luminosity functions of galaxies at $7\lesssim z \lesssim12$
(\citealt{2013ApJ...768..196S}; \citealt{2013MNRAS.432.2696M}; \citealt{2013ApJ...763L...7E}).
Recently, J. Lotz et al. have started the new deep observation program, 
Hubble Frontier Fields (HFF), that carry out deep ACS and WFC3-IR imaging
in six massive clusters. By using gravitational lensing magnifications, 
the HFF project finds faint high-redshift galaxies, behind these massive clusters, 
whose intrinsic fluxes reach $\gtrsim30$ magnitude \citep{2015ApJ...800...84C}.
Various properties of high-redshift galaxies in the HFF data are investigated and reported by many groups
\citep{2015ApJ...800...18A,2014ApJ...795...93Z,2014arXiv1409.1228O,2015ApJ...799...12I,2015ApJ...804..103K,2015MNRAS.450.3032M}.
Specifically, the studies of \citet{2015ApJ...799...12I} and \citet{2015ApJ...804..103K} are closely
related to this work.

Three independent studies of HFF have found 
an overdensity region of $z\sim8$ dropout galaxies
in the Abell 2744 cluster field (A2744C;
\citealt{2014ApJ...795...93Z,2015ApJ...800...18A,2015ApJ...799...12I}). 
\citet{2014A&A...562L...8L} investigate the HST and Spitzer band photometry of a bright source in this region.
We refer to this overdensity region as A2744z8OD.
A2744z8OD is composed of 8 galaxies at $z\sim8$ within a circular radius of $\sim6''$,
which corresponds to $\sim 300$ comoving kpc in projection.
This is an overdensity region more dense and compact than the previously identified 
ones at high redshift. For example, Brightest-of-Reionizing Galaxies (BoRG) survey
reports an overdensity of 5 galaxies at $z\sim8$ within a moderately large area of the $127'' \times 135''$ sky
\citep{2012ApJ...746...55T}. Moreover, overdensities at $z\sim 2-7$, so far studied, are
all defined by large areas with a few comoving Mpc$^2$ (\citealt{2014ApJ...792...15T}; \citealt{2007A&A...461..823V}; 
see Table 5 of \citealt{2013ApJ...779..127C}), because these previous studies aim to
target progenitors of today's massive clusters.
A2744z8OD is probably unprecedentedly compact and high-dense structure at
the redshift frontier of $z\sim8$. However, no studies have investigated 
the basic quantitative properties for A2744z8OD including galaxy surface overdensity.
A galaxy surface overdensity $\delta$ is defined by 
\begin{eqnarray}
		\delta = \frac{n - \bar{n}}{\bar{n}}, \label{eq:delta}
\end{eqnarray}
where $n$ ($\bar{n}$) is the (average) number of galaxies found in an area of the sky.
In this paper, we evaluate the observational properties of A2744z8OD quantitatively
for the first time, and discuss the physical origin of A2744z8OD via the comparisons
with the structure formation simulations of Lambda cold dark matter ($\Lambda$CDM) 
supplemented with galaxy formation models.

This paper is organized as follows. In Section \ref{sec:Data}, we present details of the observational data. 
Sample selection methods are presented in Section \ref{sec:Samples},
  and the observational properties of A2744z8OD are described in Section \ref{sec:Properties}.
  We compare these properties with the theoretical models 
  to identify the model counterpart of A2744z8OD, and to discuss 
  the physical origin in Section \ref{sec:model_comparison}.
  Finally, our results are summarized in Section \ref{sec:Summary}.
Throughout this paper, we adopt a cosmology with $\Omega_m = 0.3$, $\Omega_\Lambda = 0.7$, and $H_0 = 70$ km s$^{-1}$ Mpc$^{-1}$.

\section{Data} \label{sec:Data}

To estimate $\delta$ of A2744z8OD, we use HST deep legacy imaging data.
We search for useful HST deep legacy imaging data, available to date, 
that are deep enough to detect $z\sim8$ galaxies as faint as the members of A2744z8OD with $\lesssim 29$ mag.
Among the HST deep legacy data, we find that 8 field data sets in HFF, HUDF, and their parallel-field regions
are useful for our study, because these data reach $\simeq 29$ mag and beyond.
These data sets for our study is presented in Figure \ref{fig:distribution} and
summarized in Table \ref{summary_of_data}.
Note that the data of BoRG and CANDELS are too shallow to evaluate $\delta$ of A2744z8OD.
In this section, we explain details of the HFF, HUDF, and their parallel-field data sets.

\subsection{HFF}
HFF project plans to observe six cluster and parallel fields with ACS and WFC3-IR.
However, these observations are underway.
In this study, we use five complete data sets available to date. 
One is Abell 2744 cluster (A2744C) data which contains the overdense region A2744z8OD.
The others are four parallel field data: Abell 2744 parallel (A2744P), 
MACSJ0416.1-2403 parallel (M0416P), MACSJ0717.5+3745 parallel (M0717P), 
and MACSJ1149.5+2223 parallel (M1149P).

We do not use the associated cluster field data of MACSJ0416.1-2403, MACSJ0717.5+3745, and MACSJ1149.5+2223 in the analysis for two reasons.
The first reason is that 
the gravitational lensing effects distort and narrow
the survey area of the cluster fields in the source plane.
The second reason is that we had not made the mass models of the three clusters yet.
Recently, \citet{2015arXiv151006400K} have provided the mass models of the four clusters and 
the $z\sim8$ galaxy catalogs in the fields. 

Each data set consists of drizzled science images of seven bands in total, 
three ACS bands, F435W ($B_{435}$), F606W ($V_{606}$), and F814W ($i_{814}$), 
and four WFC3-IR bands, F105W ($Y_{105}$), F125W ($J_{125}$), F140W ($JH_{140}$), and F160W ($H_{160}$).
The $5\sigma$ limiting magnitudes are $28.8-29.1$ in $H_{160}$ band, as summarized in Table \ref{summary_of_data}.

From September 2013 to February 2014, 
Frontier Field Spitzer program (PI: T. Soifer) has obtained deep IRAC images at wavelengths of $3.6$ and $4.5 \mathrm{\mu m}$.
This program releases corrected Basic Calibrated Data (cBCD) images.
We use the Spitzer Science Center reduction software {\tt MOPEX} for the reduction of the images.
First, we run the overlap module, which performs background matching. 
Then we execute the mosaic module with a drizzled factor of 0.65.
We obtain the final mosaics with a pixel scale of $0''.6$.
The $5\sigma$ limiting magnitudes defined with a $2''.0$-diameter aperture are $24.7$ and $24.9$ in the 
3.6 and 4.5 $\mathrm{\mu m}$ bands, respectively.

\subsection{HUDF}
The HUDF is observed by HUDF (PI: Beckwith), UDF05 (PI: Stiavelli), UDF09 (PI: Illingworth), and UDF12 (PI: Ellis) programs.
Combining these data sets, the HUDF achieves the quite long exposure time especially in $Y_{105}$ band, 
which allows us to make a reliable sample of faint galaxies at $z\sim8$. 
HUDF program has conducted ultra deep observations with nine bands:
F775W and F850W in addition to the seven ACS and WFC3-IR bands of the HFF survey. 
We also use the data of two HUDF parallel fields, HUDFP1 and HUDFP2. 
These fields are observed with the filters F606W, F775W, F850W, F105W, F125W, and F160W.
The $H_{160}$-band $5\sigma$ limiting magnitudes are $29.9$ in HUDF and $29.0 - 29.2$ in HUDFP1 and HUDFP2 (Table \ref{summary_of_data}).

\setlength{\tabcolsep}{4pt}
\begin{deluxetable}{lccc}
\tablecaption{Summary of the HFF, HUDF and the Parallel Data\label{summary_of_data}}
\tablewidth{0pt}
\tablehead{
		\colhead{Field} & \colhead{Depth ($H_{160}$)\tablenotemark{a}} & \colhead{Area\tablenotemark{b}}  & \colhead{Number of $z\sim8$ Galaxies}\\
		\colhead{} & \colhead{($5\sigma$ limit mag)} & \colhead{arcmin$^2$} & \colhead{(same magnitude limit)}  
}
\startdata
A2744C	&	28.90	&	1.59\tablenotemark{c}	&	11	\\
A2744P	&	28.75	&	3.60	&	4	\\
M0416P	&	28.98	&	3.81	&	8	\\
M0717P	&	28.93	&	3.72	&	6	\\
M1149P	&	29.07	&	3.96	&	7	\\
HUDF	&	29.93	&	3.74	&	7	\\
HUDFP1	&	29.02	&	3.72	&	5	\\
HUDFP2	&	29.21	&	3.04	&	7	
\enddata
\tablenotetext{a}{The depths are defined with a $0\farcs35$-diameter aperture.}
\tablenotetext{b}{The area of each image is reduced by the conservative masking for the areas
near the bright foreground stars/galaxies and the regions with high noise levels.}
\tablenotetext{c}{The survey area in the source plane. The area in the image plane is 4.79 arcmin$^2$.}
\end{deluxetable}

\section{Samples} \label{sec:Samples}

We select $z\sim8$ dropout galaxies based on the Lyman break technique.
The selection methods are basically the same as those of \citet{2015ApJ...799...12I}
(see also \citealt{2015arXiv151107873H} for the galaxy selection in the HUDF, HUDFP1, and HUDFP2 fields).
Although Section 3 of \citet{2015ApJ...799...12I} details the methods, we briefly
explain them here. 

First, we homogenize the PSF of the images in each band in order to accurately compare 
the magnitudes of objects between the different bands. 
We convolve our data with Gaussian kernels to match the FWHMs of stars detected
in the WFC3 multi-band images. To check the reliability of this method, we obtain
flux growth curves of the stars that are total fluxes as a function of 
aperture radius in the WFC3 multi-band images. These growth curves agree
within $<10$\% in the WFC3 multi-band images, and the differences of the PSF shapes
give negligibly small impacts on our color measurements. 
The shapes of PSFs in the ACS images 
are not important, because the ACS images are not used for color measurements, but only for flux upper 
limits, in this study.
Then we make detection images 
combining $J_{125}$, $JH_{140}$, and $H_{160}$ ($J_{125}$ and $H_{160}$) images for HFF and HUDF (HUDFP1 and HUDFP2).
We run SExtractor \citep{1996A&AS..117..393B} in dual-image mode and make a source catalog in each field.
To measure colors, we use {\tt MAG\_APER} values obtained with a $0\farcs35$-diameter circular aperture.
The total magnitudes are estimated from the
{\tt MAG\_APER} values and the aperture correction 0.82 mag.
These total magnitudes agree well with the {\tt MAG\_AUTO} magnitudes
(see Section 3.1 of \citealt{2015ApJ...799...12I}).
 
\begin{figure*}
	\plotone{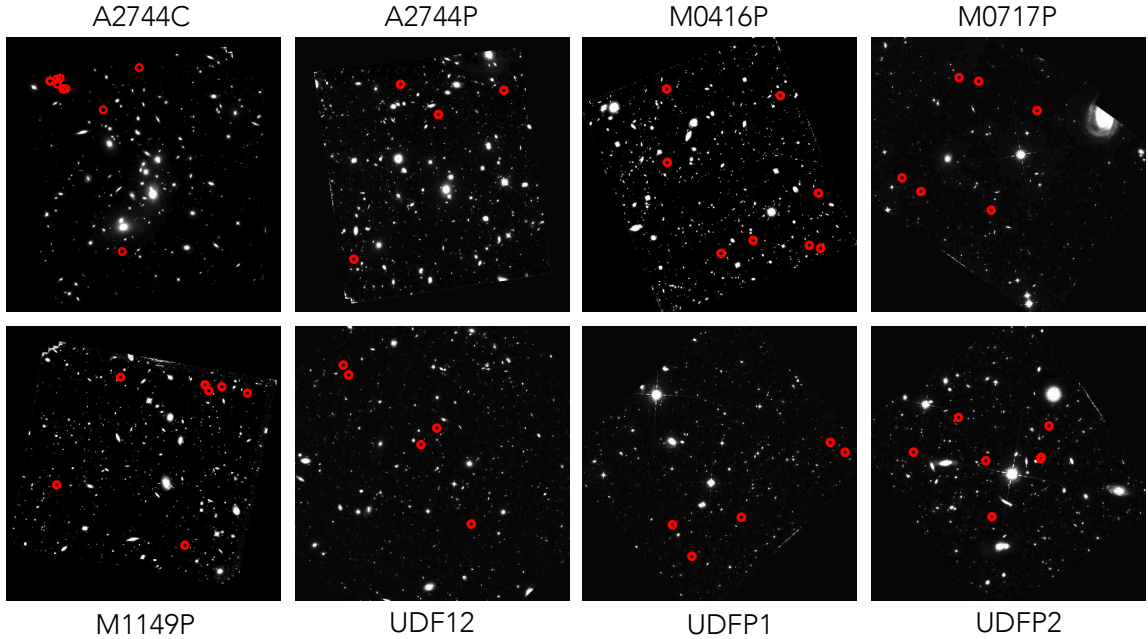}
	\caption{$H_{160}$ images of the 8 fields that we use. The red circles represent the positions of $z\sim8$ galaxies. 
	}
	\label{fig:distribution}
\end{figure*}

The selection criteria for $z\sim8$ galaxies are represented by following equations:
\begin{eqnarray}
		Y_{105} - J_{125} > 0.5 \\
		J_{125} - H_{160} < 0.4. 
\end{eqnarray}
that is the same as the one used in \citet{2015ApJ...799...12I} (see
\citealt{2013ApJ...768..196S} for the original definition).
To obtain a homogeneous $z\sim 8$ galaxy sample from the different fields,
we apply these selection criteria to the source catalogs of all fields, and remove
sources whose optical band magnitudes are brighter than the $2\sigma$ detection levels.
We also apply the magnitude limits of $J_{125}\simeq 29.3$ and $H_{160}\simeq 29.1$ magnitudes
in a $0\farcs35$-diameter circular aperture
that correspond to 
the $3.5\sigma$ significance levels in the shallowest data,
and identify 57 galaxies in total whose positions are indicated in Figure \ref{fig:distribution}. 
We have checked the consistency of our catalog with the other published catalogs.
Our catalog of A2744C is consistent with those in \citet{2014ApJ...795...93Z}, \citet{2015ApJ...800...18A}, and \citet{2015ApJ...800...84C}.
In HUDF, we reproduce sources in the catalog of \citet{2013MNRAS.432.2696M} and \citet{2013ApJ...768..196S}
down to the limiting magnitudes of our study. 
Although the source spatial distribution in M1149P (Figure \ref{fig:distribution}) appears 
to be biased to the edge of the image,
we confirm that these are reliable sources in the securely deep imaging area by visual inspection.

\begin{figure}
	\epsscale{1.3}
	\plotone{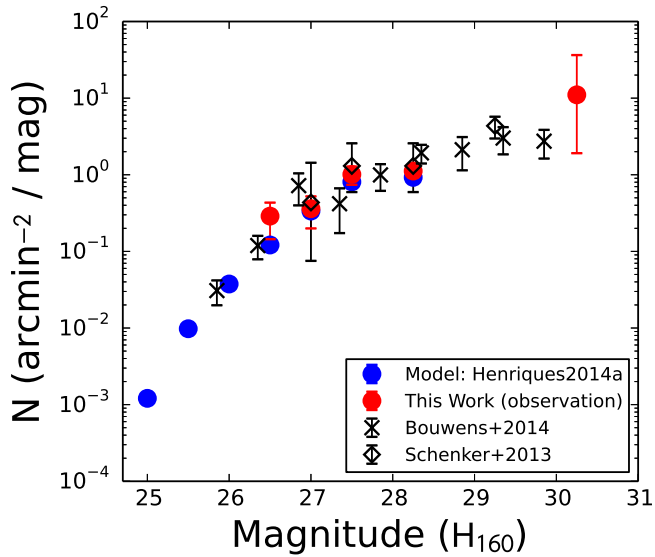}
	\caption{Surface number densites of $z\sim8$ galaxies as a function of total $H_{160}$ magnitude.
			The red circles denote the surface number densities of our $z\sim 8$ galaxy sample that are
			corrected for the lensing effects of the magnification and the distortion in the image plane.  
			We show the surface number densities of model $z\sim8$ galaxies (Section \ref{sec:Simulation}) with the blue circles,
			and those of previous observational data with the crosses and triangles.
	}
	\label{fig:surface_density}
\end{figure}

\begin{figure*}
	\plotone{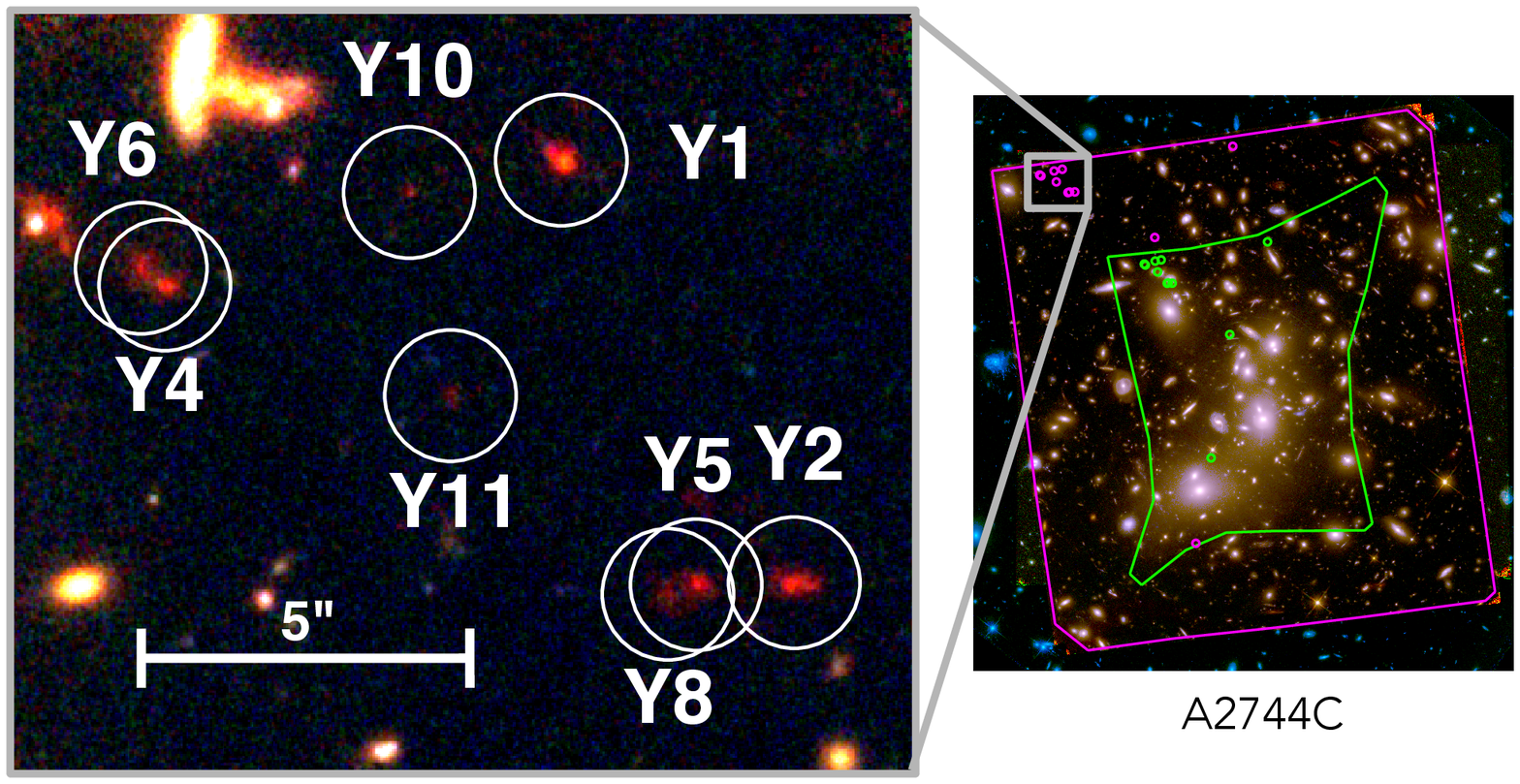}
	\caption{
			{\it Right:} False color image of the A2744C field. The magenta line and circles represent the boundary of the used data and the positions of dropout galaxies at $z\sim8$, respectively, in the image plane. The green line and circles are the same, but in the source plane. 
		{\it Left:} The magnified false color image in the A2744z8OD region. The white circles indicate the positions of the 8 $z\sim 8$ galaxies forming A2744z8OD.
	}
	\label{fig:A2744Cz8OD}
\end{figure*}

Figure \ref{fig:surface_density} presents the surface number densities of these galaxies, and indicates that the surface number densities
are consistent with those of previous studies. 
The total magnitudes are corrected for the magnification by the gravitational lensing effects.
Here we use the mass model of \citet{2015ApJ...799...12I} which is constructed 
with the parametric gravitational lensing package {\tt GLAFIC} \citep{2010PASJ...62.1017O}.
We use magnification-corrected magnitudes for the lensed sources in this paper.
By these source detection and selection methods applied to all of our data sets,
we again find the overdensity of $z\sim 8$ galaxies, A2744z8OD, in the north east region of A2744C (Figure \ref{fig:A2744Cz8OD}), 
as reported by previous studies (\citealt{2014ApJ...795...93Z}; \citealt{2015ApJ...800...18A}; \citealt{2015ApJ...799...12I}). 
A2744z8OD consists of eight $z\sim 8$ galaxies within a $6''$-radius circle,
which are dubbed ID Y1, Y2, Y4, Y5, Y6, Y8, Y10, and Y11 in \citealt{2015ApJ...799...12I}.
The magnification factors of these galaxies are $\sim 1.5$ \citep{2015ApJ...799...12I}.
Interestingly, Figure \ref{fig:distribution} shows no overdense regions 
similar to A2744z8OD in the rest of our fields.
We confirm that A2744z8OD is a rare overdensity.

\section{Observational Properties of A2744z8OD} \label{sec:Properties}

\subsection{Overdensity}
\begin{figure}
	\epsscale{1.3}
	\plotone{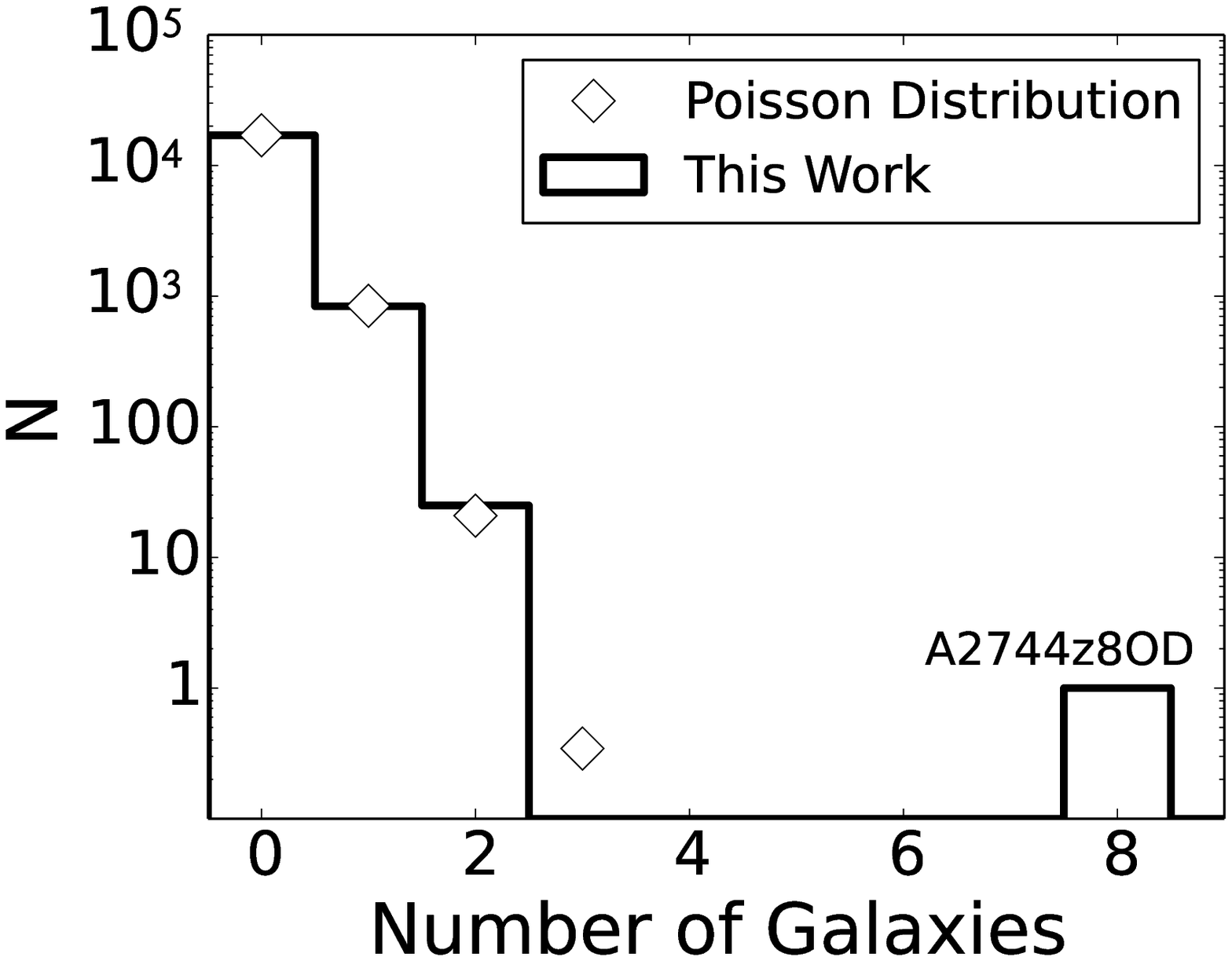}
	\caption{
			Galaxy-number distribution for $z\sim8$ galaxies in A2744z8OD and the other positions of A2744C, 
			A2744P, M0416P, M0717P, HUDF, HUDFP1, and HUDFP2 fields (solid histogram). 
			The excess at 8 in the number of galaxies corresponds to A2744z8OD.
			The diamonds indicate a Poisson distribution whose average number density is the same as our observations.
	}
	\label{fig:overdensity}
\end{figure}

To estimate $\delta$ of A2744z8OD, we count the $z\sim 8$ galaxy numbers within a
$6''$-radius ($\simeq 30$ physical kpc) circle that corresponds to the angular size of A2744z8OD. 
About 16,000 circles are homogeneously placed in our 8 fields as well as
the area centered at A2744z8OD. In the A2744C field, we use the mass model of 
\citet{2015ApJ...799...12I} to correct for the survey area distortion, and measure 
the galaxy numbers in the source plane.
Figure \ref{fig:overdensity} presents
the histogram of the galaxy numbers.

The galaxy-number distribution is basically explained by a Poisson distribution shown with the diamonds in Figure \ref{fig:overdensity}.
However, the galaxy number in A2744z8OD which contains eight galaxies 
significantly exceeds the Poisson distribution.
If the galaxies are randomly distributed, the probability to find the A2744z8OD excess is only $1.9\times10^{-11}$
in the Poisson statistics. It suggests that the galaxy overdensity of A2744z8OD is not a statistical fluctuation, but a physical structure.

\begin{figure}
	\epsscale{1.2}
	\plotone{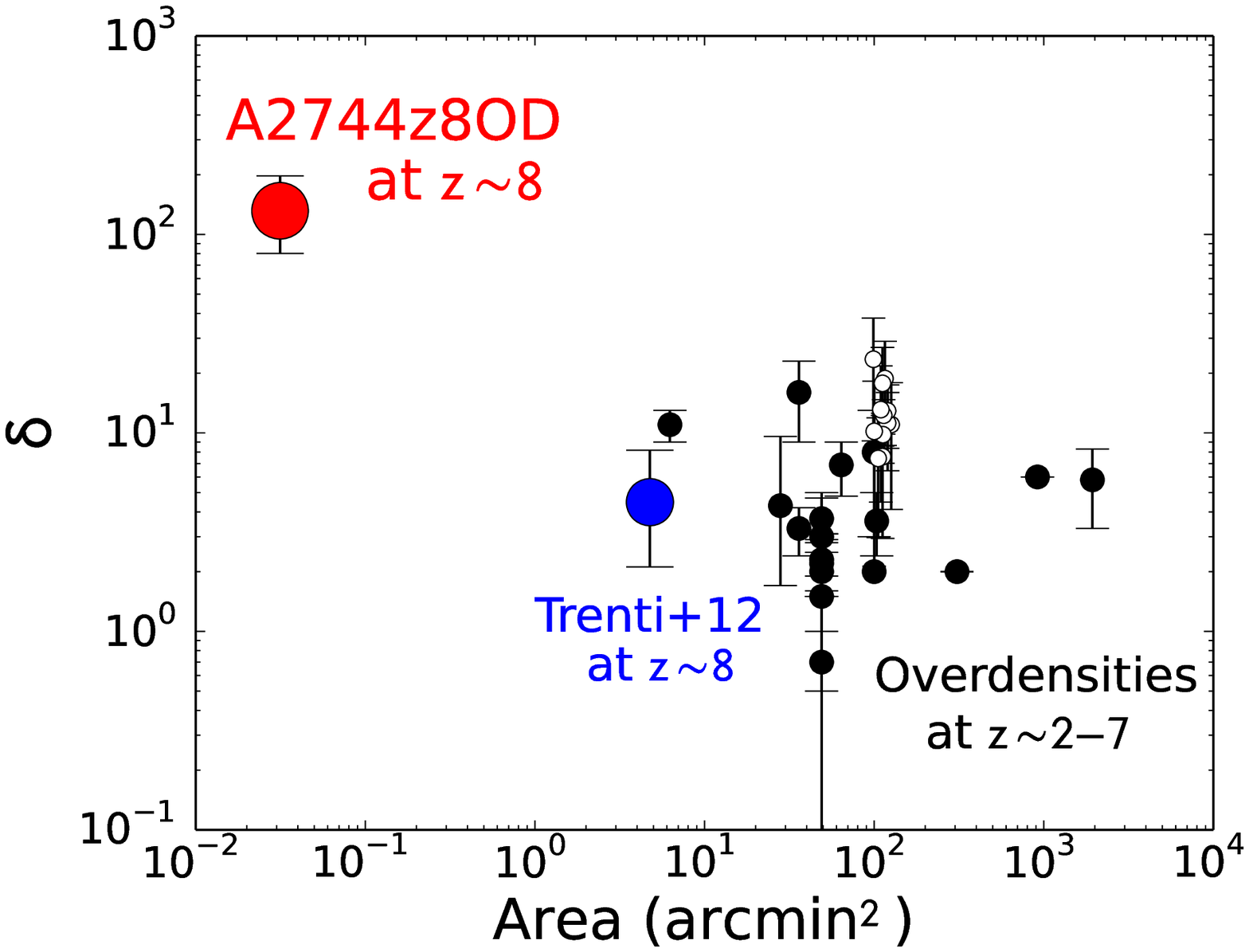}
	\caption{Galaxy surface overdensity $\delta$ as a function of area.
	The area is the one used for the $\delta$ measurements.
	The red circle represents A2744z8OD at $z\sim 8$, while the blue circle denotes
	the overdensity at $z\sim 8$ reported by \citet{2012ApJ...746...55T}.
	The black filled and open circles indicate the overdensities at $z\sim 2-7$
	summarized in Table 5 of \citet{2013ApJ...779..127C} 
	and Table 1 of \citet{2015arXiv151204956F}, respectively.
	}
	\label{fig:chiang2013}
\end{figure}

We derive $\delta$ of A2744z8OD with Equation (\ref{eq:delta}). 
Following the procedures above, we use an area of a $6''$-radius circle
to define galaxy numbers for $\delta$.
By these procedures, we obtain $\delta=132^{+66}_{-51}$ for A2744z8OD.
Figure \ref{fig:chiang2013} compares the $\delta$ value of A2744z8OD and 
those of overdensities at $z\sim 2-8$ reported in previous studies.
Here we use the summary of the overdensities at $z\sim 2-7$ presented 
in \citet{2013ApJ...779..127C} and \citet{2015arXiv151204956F}, and the recent report of the overdensity at $z\sim 8$ 
\citep{2012ApJ...746...55T}. Because $\delta$ depends on the area used for
galaxy number counts, Figure \ref{fig:chiang2013} shows $\delta$ as a function of area.
Clearly, A2744z8OD is placed at a position very different from the others
(Figure \ref{fig:chiang2013}). A2744z8OD is the most compact and dense high-$z$ structure,
so far, identified.

There are three  possible sources of uncertainties in the $\delta$ estimate.
First, 
there is a possibility that close pair sources, Y5-Y8 and Y4-Y6, are not two different
galaxies, but single galaxies with clumpy structures. The distance between
Y5 and Y8 is about $0.5$ arcsec, corresponding to $\sim 2.8$ kpc projected on the sky,
that is the smallest among the close pairs.
The typical size of $z\sim8$ galaxies is about $0.5$ kpc \citep{2015ApJS..219...15S}
that is smaller than the Y5-Y8 galaxy distance ($\sim 2.8$ kpc) by a factor of 5-6.
Some recent studies show that the size of high-redshift galaxies at the bright end is $\sim1-2$ kpc (\citealt{2013ApJ...777..155O}; \citealt{2016MNRAS.457..440C}),
which is still smaller than the Y5-Y8 galaxy distance.
Thus, it is unlikely that the close pairs including Y5-Y8 pair are clumps in single galaxies.
Second, $z\sim 8$ galaxies can be blended or obscured by foreground sources, and the
detection completeness of $z\sim 8$ galaxies may be reduced.
We calculate the area that foreground sources occupy,
and find that it is about 4\% of the total survey area.
By the detection incompleteness, the value of $\delta$ would change by 4\%
that is negligibly small.
The third source of the uncertainty is the foreground source contamination in our $z\sim 8$ galaxy sample.
We discuss effects of the foreground source contamination in Section \ref{sec:photoz}.

\subsection{Photometric Redshifts}\label{sec:photoz}
\begin{figure*}
	\epsscale{1.2}
	\plotone{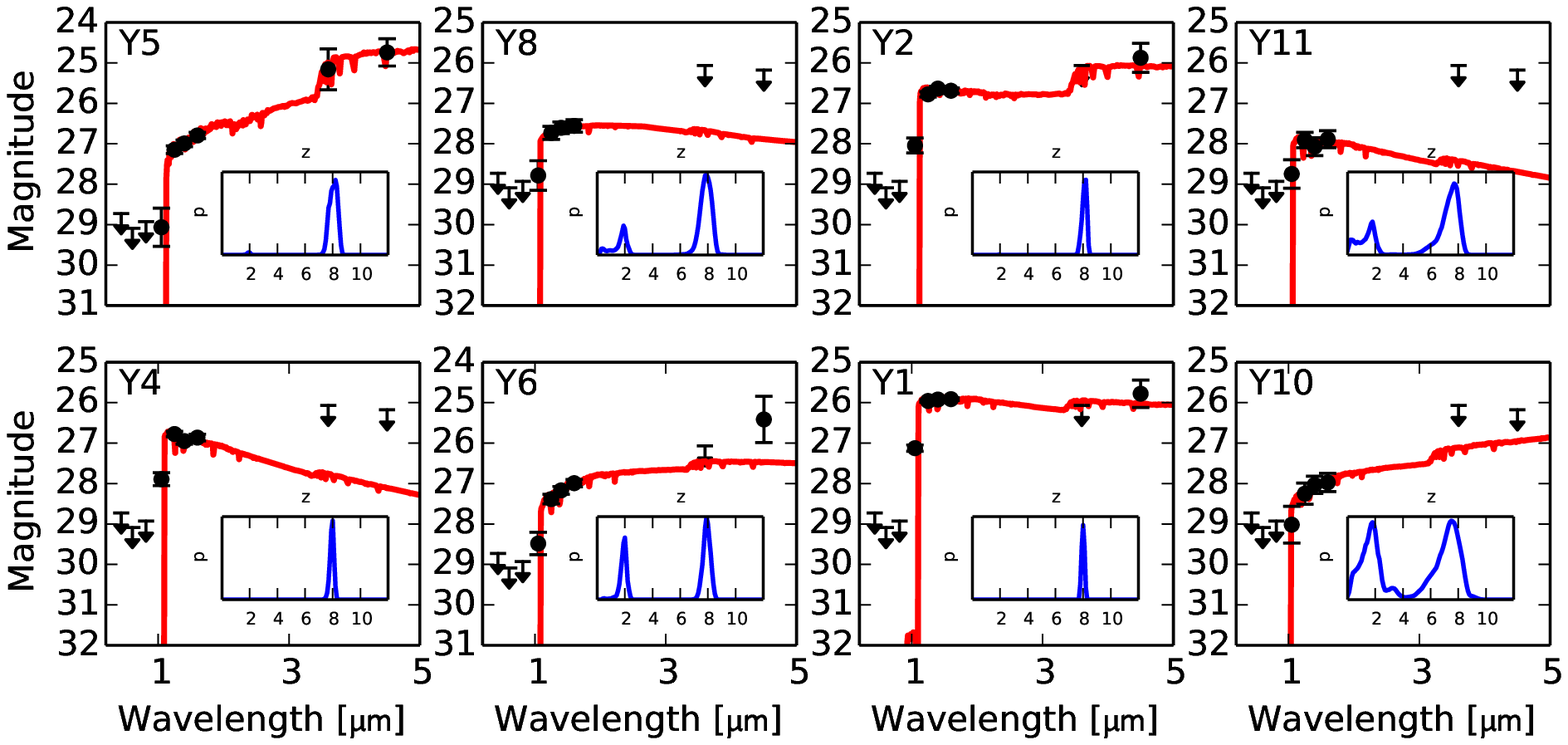}
	\caption{
			The SED fits of the eight galaxies in the overdense region.
			Black arrows present the $2\sigma$ upper limits.
			We also show the redshift probability distributions in each panel.
	}
	\label{fig:spec_pdf}
\end{figure*}

To derive the contamination rate of foreground sources, we compute the photometric redshifts of our eight A2744z8OD galaxies.
Because rest-frame optical photometry is key for determining photometric redshifts,
we measure the source fluxes in IRAC $3.6$ and $4.5 \mathrm{\mu m}$ bands.
Because the IRAC images have a spatial resolution worse than the HST images,
we model the IRAC images with the public software for the photometry of low resolution images, {\tt T-PHOT} \citep{2015A&A...582A..15M}.
We combine $J_{125}$, $JH_{140}$, and $H_{160}$ images, and use the combined HST image for the high resolution reference image of {\tt T-PHOT}. 
The source catalog of the reference image is made with SExtractor.
{\tt T-PHOT} convolves the reference image with the PSF of IRAC,
and the convolved reference image is fitted to the real IRAC image
for each source. We thus obtain the best-fit IRAC source fluxes.
Figure \ref{fig:spec_pdf} shows the IRAC magnitudes of the eight A2744z8OD galaxies, 
together with the ACS and WFC3 magnitudes.

We calculate the photometric redshifts with a new version of the code Hyperz, New-Hyperz,
which is originally developped by \citet{2000A&A...363..476B}.
We assume the SEDs of \citet{2003MNRAS.344.1000B} models,
Chabrier IMF \citep{2003PASP..115..763C}, exponential SFR ($\tau = 0.01 - 3$ Gyr),
metallicity ($Z/Z_\odot = 0.02 - 1.0$), Calzetti dust attenuation ($A_V = 0.0 - 1.5$; \citealt{2000ApJ...533..682C}).
Figure \ref{fig:spec_pdf} present the best-fit SEDs and the redshift probability distributions.
All the eight A2744z8OD galaxies have the best-fit SEDs at $z\sim8$.
Although some faint galaxies have moderately high low-redshift contamination rates, 
the total contamination rate is only 16\%. This value is consistent with the estimate of 
the contamination rate in \citet{2015ApJ...799...12I}.
This uncertainty is added to the error of $\delta$.

\subsection{Stellar Mass}
\label{sec:stellar_mass}
\begin{figure}
	\epsscale{1.3}
	\plotone{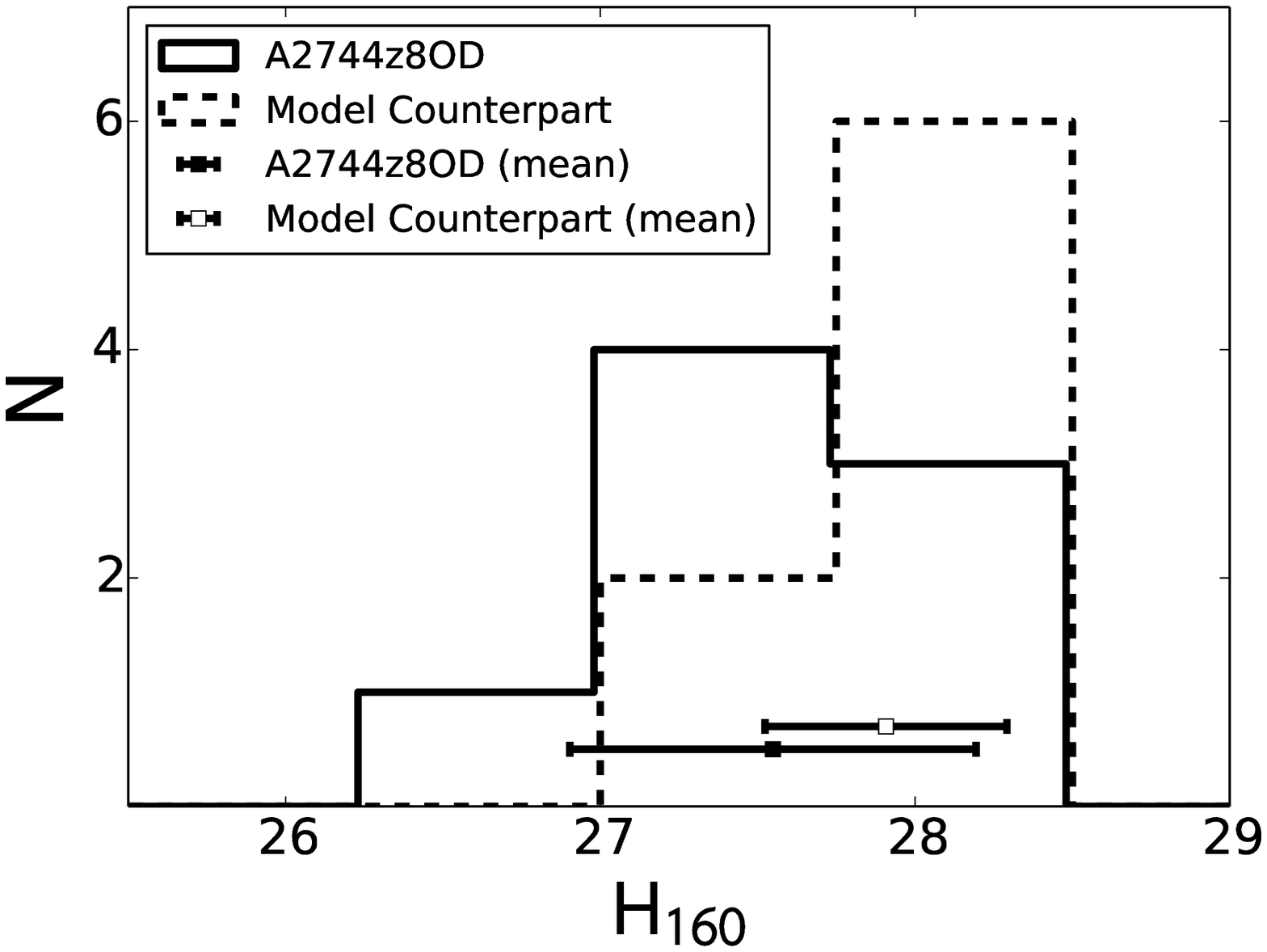}
	\caption{
			$H_{160}$-band total magnitude histograms of $z\sim8$ galaxies in A2744z8OD (solid line) and 
			the model counterpart Modelz8OD (dashed line). 
			The black filled and open squares indicate the mean magnitudes of the A2744z8OD and Modelz8OD sources, respectively.
			The associated error bars denote the deviation values of the magnitude distribution.
	}
	\label{fig:UV_hist}
\end{figure}

We estimate the stellar mass of A2744z8OD with the mass-to-luminosity relation 
in \citet{2013ApJ...777L..19L} that is derived by SED fitting of $z\sim 8$ dropout galaxies.
Here we assume \citet{2003PASP..115..763C} IMF in our study to obtain stellar masses
that can be compared with the models shown in the next section.
Because \citet{2013ApJ...777L..19L} use \citet{1955ApJ...121..161S} initial mass function (IMF),
we divide the mass-to-luminosity relation of \citet{2013ApJ...777L..19L} by a factor of 1.8:

\begin{eqnarray}
		M_\ast = 0.9^{+0.6}_{-0.4}/1.8 \times 10^9 M_\odot \ \mathrm{for}\ H\simeq 27.3
	\label{eq:stellar_mass}
\end{eqnarray}
We confirm that this relation is consistent with the results of the recent study \citep{2015arXiv150705636S}.
Figure \ref{fig:UV_hist} presents the histogram of the rest-frame UV ($H_{160}$ band) total magnitudes for the 8 galaxies in A2744z8OD.
We calculate the total UV luminosity of these 8 galaxies, and calculate the total stellar mass with the total UV luminosity
and the Equation (\ref{eq:stellar_mass}). The total stellar mass is $(4\pm2) \times 10^9 M_\odot$.
Note that this total stellar mass is comparable to the one of today's Large Magellanic Cloud (Figure 3 of \citealt{2015arXiv150700676D}).
We summarize the properties of A2744z8OD in Table \ref{property_of_OD}.

\section{Model Comparison and Discussion}\label{sec:model_comparison}

\subsection{Model}
\label{sec:model}

We investigate theoretical models whether there exist a compact and dense overdensity at $z\sim 8$ 
similar to A2744z8OD in galaxy formation models based on the $\Lambda$CDM structure formation framework.
We use Henriques2014a galaxy formation models \citep{2015MNRAS.451.2663H}
made of Millennium \citep{2005Natur.435..629S} and Millennium-II \citep{2009MNRAS.398.1150B} Simulation
for the structure formation and the semi-analytic recipes for the baryonic processes.
Henriques2014a models include 24 light-cone outputs. Each light-cone has a 3.14 degree$^2$ area.
The light-cone outputs provide catalogs with magnitudes of the ACS and WFC3-IR filter system.

\subsection{Model Counterpart of A2744z8OD}
\label{sec:Simulation}

\begin{figure}
	\epsscale{1.2}
	\plotone{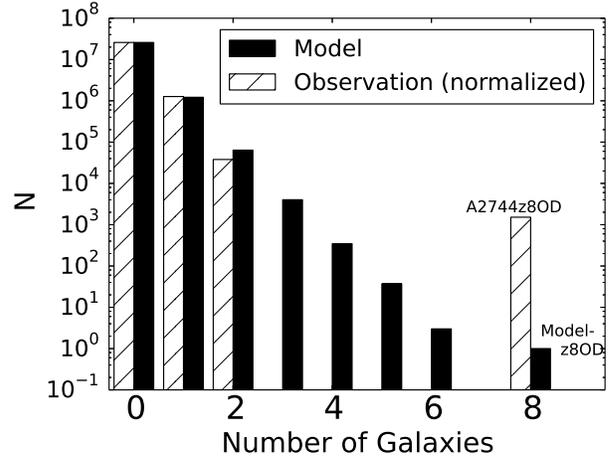}
	\caption{Galaxy-number distribution of $z\sim8$ galaxies in the model catalogs (black histogram).
			For comparison, we also show the galaxy-number distribution from the observations with dashed histograms.
			The dashed histograms are the same as those in Figure \ref{fig:overdensity}, but 
			the total galaxy numbers of the observations are scaled to match those of the models. 
			Note that the survey area of the simulations is much larger than that of the observations.
	}
	\label{fig:delta_distribution}
\end{figure}

Using the catalogs of the light-cones, we select $z\sim 8$ dropout galaxies with the color criteria
same as our observations (Section \ref{sec:Samples}).
Figure \ref{fig:surface_density} presents the surface number densities 
of the model dropout galaxies at $z\sim 8$.
The surface number densities of the models are consistent with those of observations.
The redshift distribution of color-selected galaxies in the simulation has the peak at $z=7.8$,
which is also similar to that in the observation.
Figure \ref{fig:delta_distribution} shows the $\delta$ distribution of
the model dropout galaxies at $z\sim 8$, and compares
with the one of the observations, which is the same as Figure \ref{fig:overdensity} but with a different scaling for comparison.
Figure \ref{fig:delta_distribution} presents that the $\delta$ distributions of the models and the observations are 
almost the same below 4 in number of galaxies.
Because the total area of the observations is significantly smaller than the one of the models,
the observations do not precisely measure the $\delta$ distribution equal to or greater than 4 in number of galaxies,
as indicated in Figure \ref{fig:overdensity}.
Thus, Figures \ref{fig:surface_density} and \ref{fig:delta_distribution} suggest that
the models well reproduce the abundance and clustering of the observed $z\sim 8$ galaxies.
 
In Figure \ref{fig:delta_distribution}, we find one region that has the $z\sim8$ galaxy number same as A2744z8OD.
It is located at $\mathrm{R.A.}=0.64333$ deg and $\mathrm{Decl.}=0.52677$ deg in the catalog Henriques2014a.cones.MRscPlanck1\_BC03\_010.
We refer this region as Modelz8OD.
Figure \ref{fig:UV_hist} presents the UV magnitude distribution of Modelz8OD.
The mean of the UV magnitude distribution of Modelz8OD is similar to the one of A2744z8OD
within the $1\sigma$ level.
We calculate total stellar mass of Modelz8OD from the UV magnitudes with 
Equation \ref{eq:stellar_mass} in the same manner as Section \ref{sec:stellar_mass},
and obtain $2.4\times 10^{9} M_\odot$. This value is comparable with the one of A2744z8OD
within a factor of 1.5 difference.
We thus confirm that Modelz8OD is similar to A2744z8OD in the UV magnitude distribution
and the total stellar mass.

Table \ref{property_of_OD} summarizes the properties of Modelz8OD.
Note that there are two total stellar masses for Modelz8OD. One is the estimate
with the Equation (\ref{eq:stellar_mass}) ($2.4\times 10^{9} M_\odot$), 
and the other is the value taken from the Henriques2014a model catalogs
($9.2 \times 10^{8} M_\odot$) that assume the same IMF of \citet{2003PASP..115..763C}.
Because the Henriques2014a model catalog values
provide accurate stellar masses of Modelz8OD, we hereafter
use the stellar mass values in the Henriques2014a model catalogs
for Modelz8OD.

\begin{figure}
	\epsscale{1.2}
	\plotone{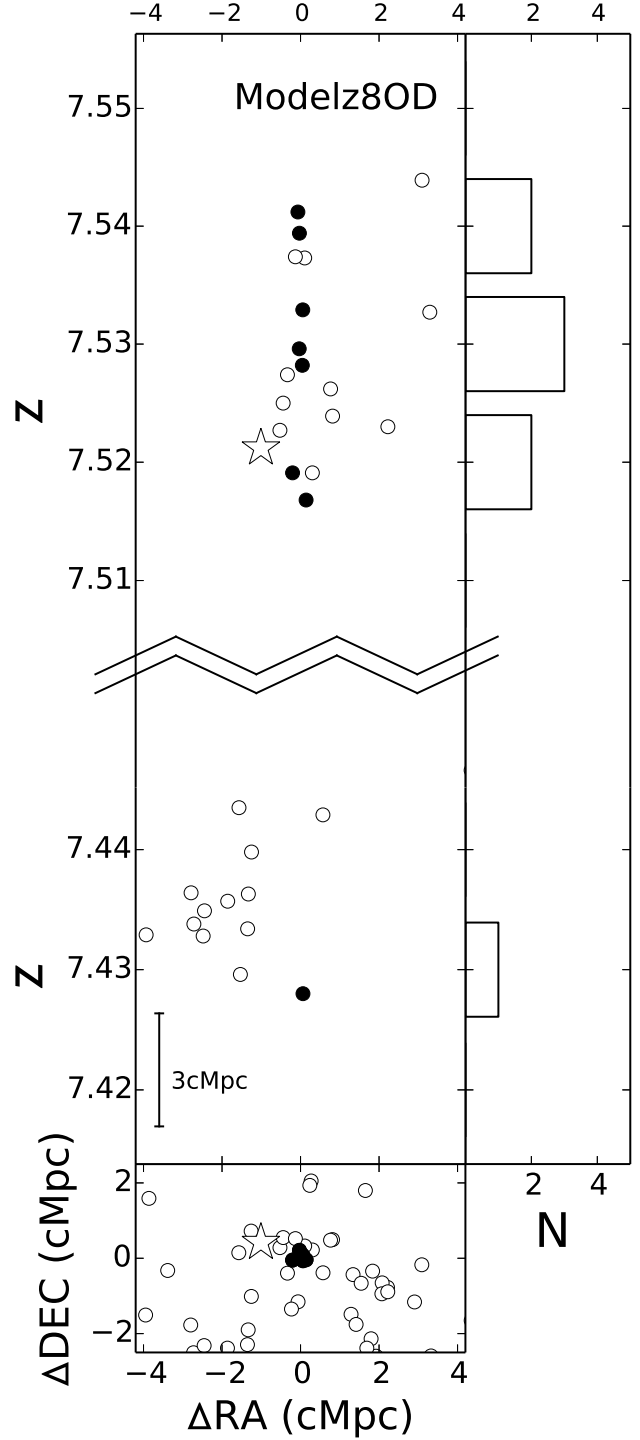}
	\caption{
			Three-demensional map of model $z\sim 8$ galaxies in and around Modelz8OD. 
			The filled (open) circles denote positions of $z\sim8$ galaxies in (out of) Modelz8OD that is
			defined by a circle of $6''$ radius on the sky.			 
			The white star represents the most massive BCG progenitor located 1 Mpc away from the overdense region.
			The top left panel presents the distribution of $z\sim8$ galaxies in transverse vs. redshift directions.
			The bottom panel shows the distribution projected on the sky.
			The top right panel is the redshift-distribution histograms for the eight galaxies of Modelz8OD.
	}
	\label{fig:display}
\end{figure}

We investigate the structure of Modelz8OD that is selected with
a circle of $6''$-radius ($\sim30$ physical kpc) in projection.
Figure \ref{fig:display} shows the three-demensional map in and around Modelz8OD.
Seven out of eight galaxies of Modelz8OD are physically close to each other
in the redshift space. The redshift-range width of these seven galaxies is $\Delta z \sim 0.03$ ($\sim 10^4$ km s$^{-1}$).
This width corresponds to $\sim 10$ comoving Mpc ($\sim 1$ physical Mpc), if these galaxies have no peculiar motions
against the Hubble flow.
Because the redshift range of the color selection is $\Delta z \sim 1$ (Section \ref{sec:Samples}),
these seven galaxies are very strongly clustered in the redshift space.
Figure \ref{fig:display} indicates that 
the distribution of the seven galaxies is largely elongated in the line of sight, and that
the seven galaxies of Modelz8OD are probably found in a filament of the large scale structure of the universe.
The high $\delta$ value of Modelz8OD is not only made by the intrinsically-high density but
by the projection effect.
If A2744z8OD is similar to the model counterpart of Modelz8OD,
most of the A2744z8OD galaxies would form a physical association in the redshift space.
Moreover, these A2744z8OD galaxies may reside at a filament of the large scale structures.

\subsection{Halo and Stellar Mass Evolution\\ 
Suggested by the Model}
\label{sec:Discussion}

\begin{figure}
	\epsscale{1.2}
	\plotone{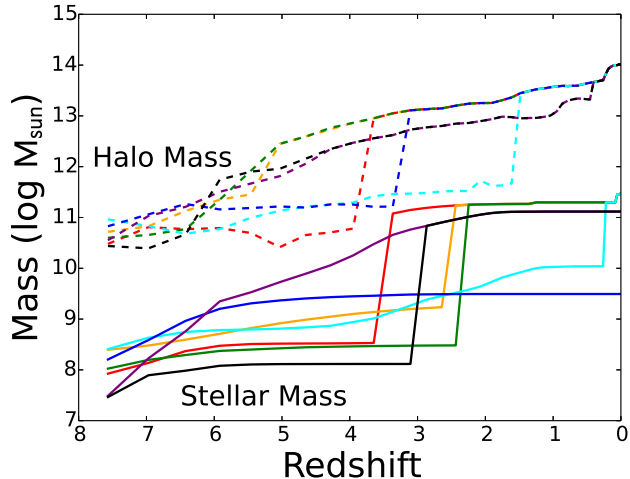}
	\caption{Evolutions of the halo masses (dashed lines) and stellar masses (solid lines) evolutions 
		for the seven galaxies in Modelz8OD. One color of dashed and solid lines corresponds to one galaxy.
		At $z<4$, four out of the seven galaxies merge into the most massive progenitor of the BCG.
	}
	\label{fig:mass_evolution}
\end{figure}

We investigate the mass evolutions of the seven galaxies of Modelz8OD that form the physical association (Section \ref{sec:Simulation})
based on Henriques2014a models that trace mass evolutions and merger histories of galaxies.
Figure \ref{fig:mass_evolution} presents the halo and stellar mass evolutions of the seven Modelz8OD galaxies. 
Figure \ref{fig:mass_evolution} indicates that the halos of the seven galaxies merge into a single massive halo 
with a mass of $\sim 10^{14} M_\odot$ at $z=0$.
Because this mass is comparable to a galaxy cluster today, 
Modelz8OD is regard as a cluster progenitor, or protocluster, 
at $z\sim 8$.

The Henriques2014a model catalog has flags indicating whether a galaxy 
is included in the central galaxy based on the friend-of-friends algorithm.
Four out of the seven galaxies merge into the central galaxy,
the brightest cluster galaxy (BCG), that is the most massive galaxy 
in this cluster system at $z=0$. 
Thus, more than a half of Modelz8OD galaxies are building blocks of
the BCG. In this sense, Modelz8OD is a progenitor of
the cluster core. Hereafter, we refer to this BCG as Modelz8OD BCG.
Note that the $z\sim8$ galaxies in the overdense region do not provide the majority of mass of the BCG,
but a part of mass of the BCG.
At $z\sim8$, the galaxies in the overdense region contribute $20\%$ of the total stellar mass of the BCG progenitors.
The most massive progenitor is located about 1 Mpc away from the overdense region at $z\sim8$ (Figure \ref{fig:display}).
Its stellar mass is $2.1 \times 10^9 M_\odot$.
Because Modelz8OD is associated with the most massive galaxy,
it is possible that A2744z8OD would be also accompanied by 
such a massive galaxy that would be located outside our survey area.
This comparison implies that such galaxy overdensities can be tracers of massive BCG progenitors.

In order to quantify the relation between overdensities and massive BCGs, 
we calculate the probability that a galaxy at $z\sim8$ becomes a BCG in a cluster with the halo mass of $> 10^{14} M_\odot$ at $z=0$. 
We select model galaxies at $z\sim8$ whose UV magnitudes are brighter than $-20$ mag, which is a typical magnitude of galaxies in A2744z8OD.
We find that the probability is $60\%$ for the galaxies in overdensities of $\delta \gtrsim 60$, and $30\%$ for the galaxies in the whole region.
The fraction of BCG progenitors in galaxies in overdensities is larger than that in the average $z\sim8$ galaxies.

\setlength{\tabcolsep}{4pt}
\begin{deluxetable*}{lcccccc}
\tablecaption{Properties of A2744z8OD and Modelz8OD\label{property_of_OD}}
\tablewidth{0pt}
\tablehead{
		\colhead{} & \colhead{N\tablenotemark{a}} & \colhead{$\delta$}  & \colhead{$M_{\ast,M/L}$\tablenotemark{b,f}} & \colhead{$M_\ast$\tablenotemark{c,f}}  & \colhead{$M_{\mathrm{halo\ z=0}}$\tablenotemark{d}} & \colhead{$M_{\mathrm{\ast\ z=0}}$\tablenotemark{e,f}}\\
		\colhead{} & \colhead{} & \colhead{} & \colhead{[$M_\odot$]} & \colhead{[$M_\odot$]} & \colhead{[$M_\odot$]} & \colhead{[$M_\odot$]}  
}
\startdata
A2744z8OD	&	$8$	&	$132^{+66}_{-51}$	&	$(4\pm2) \times 10^9$	& \nodata	&	\nodata	&	\nodata	\\
Modelz8OD &	$8$	&	$159$	&	$2.4 \times 10^9$ & $9.2 \times 10^8$	&	$9.4 \times 10^{13}$	&	$3.0 \times 10^{11}$	
\enddata
\tablenotetext{a}{Galaxy number of $z\sim 8$ galaxies in a $6''$-radius circle.}
\tablenotetext{b}{Stellar mass derived from the mass-to-luminosity relation (Equation \ref{eq:stellar_mass}).}
\tablenotetext{c}{Stellar mass given by the Henriques2014a model catalogs.}
\tablenotetext{d}{$z=0$ halo mass taken from the Henriques2014a model catalog.}
\tablenotetext{e}{$z=0$ stellar mass taken from the Henriques2014a model catalog.}
\tablenotetext{f}{\citet{2003PASP..115..763C} IMF.}
\end{deluxetable*}

\begin{figure}
	\epsscale{1.2}
	\plotone{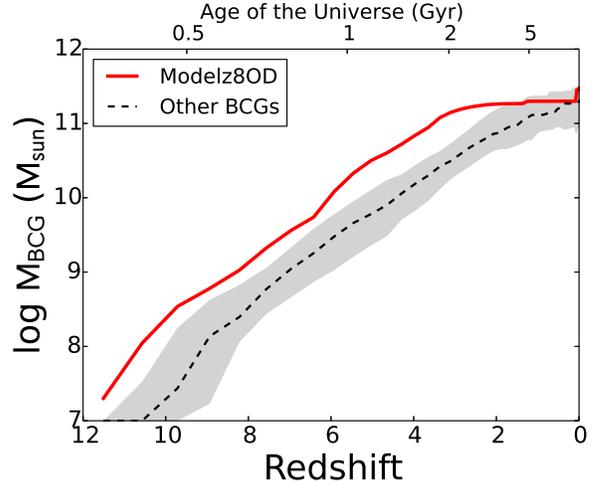}
	\caption{
			Stellar mass evolutions of BCGs in the Henriques2014a models.
			The red line shows the stellar mass evolution of the Modelz8OD BCG.
			Note that Modelz8OD BCG is not associated with Modelz8OD at high redshifts.
			The black dashed line denotes the median of stellar masses of the other BCGs which 
			reside in clusters with a halo mass of $\sim 10^{14} M_\odot$ at $z=0$.
			The gray region represents the 16th-84th percentiles of the BCG stellar-mass distribution.
	}
	\label{fig:stellar_evolution}
\end{figure}

Finally, we compare the stellar-mass evolution of Modelz8OD BCG with those of the other BCGs in the models.
We select BCGs in clusters whose halo masses are $\sim 10^{14} M_\odot$ at $z=0$ that is comparable with the one of Modelz8OD BCG.
Figure \ref{fig:stellar_evolution} presents the stellar-mass evolutions of these BCGs.
At $z\sim8$, the Modelz8OD BCG is more massive than the other BCGs at the $\sim 2\sigma$ level.
Modelz8OD BCG is one of the most stellar massive, i.e. matured, systems among the BCG progenitors at $z\sim 8$.
We investigate the stellar-mass evolution of the BCGs to higher and lower redshifts in Figure \ref{fig:stellar_evolution}.
Towards low redshift, the stellar masses of the other BCGs catch up with 
the one of Modelz8OD BCG at $z\lesssim2$.
Because BCGs assemble their masses from a volume larger than the $6''$-radius circle at $z\sim 8$,
the mass growths of some high-$z$ compact+dense systems such as Modelz8OD could slow down
at $z\lesssim2$. 
This indicates that the dense systems at $z\sim 8$ 
are not necessarily the progenitors of the most massive systems at $z\lesssim2$.
Evolution to the cluster-sized systems depends on the overdensities defined
by a circle radius larger than $\sim 6''$ at $z\sim 8$.
Towards high redshift, Modelz8OD BCG is significantly more massive than the other BCGs at least
up to $z\sim 12$.
Although the models have a mass resolution of $\sim10^{7} M_\odot$ that do not allow us 
to trace back to $z\gtrsim 12$, the massive system of Modelz8OD BCG should start 
forming at a very early epoch.
Because $\delta$ is defined by a very small circle of $6''$ radius (0.3 comoving Mpc) at $z\sim 8$ 
that is much smaller than the cluster-scale fluctuations,
large $\delta$ systems like Modelz8OD are progenitors of present-day massive galaxies 
that start forming earlier ($z>12$) than the other massive galaxies.
These results suggest a possibility that overdensities defined by such a small-radius circle 
allow us to identify the site of the earliest star-formation took place in the universe.
If A2744z8OD is truly similar to the model counterpart of Modelz8OD,
the system of A2744z8OD would become a cluster today. In this process, some of the A2744z8OD member
galaxies could merge into a BCG to form a core of the cluster. 
We may witness the early cluster core
formation with A2744z8OD.

These model comparisons provide an interesting insight into 
future high-$z$ galaxy overdensity studies.
Wide-field observation programs such as Subaru/Hyper Suprime-Cam surveys
will probably reveal a number of compact dense systems at high-$z$, which will complement
popular studies of protoclusters \citep{2013ApJ...779..127C,2014ApJ...792...15T}
defined by overdensities in a large-radius circle.

\section{Summary}
\label{sec:Summary}

We investigate the overdense region of eight dropout galaxies at $z\sim8$, A2744z8OD, behind Abell 2744 cluster,
that is originally pinpointed by the several HFF studies. This work identifies, for the first time, 
the A2744z8OD's physical properties of very compact and dense overdensity quantitatively
that are clearly different from those reported in previous studies of galaxy overdensities at $z\gtrsim 2$.
The detailed comparisons with theoretical models are made to understand the physical origin of 
A2744z8OD.
The results of this study are summarized below.

\begin{enumerate}
\item We make the homogeneous $z\sim 8$ galaxy sample with the eight field data of Hubble legacy images 
that are deep enough to detect galaxies as faint as the members of A2744z8OD with $\lesssim 29$ mag,
and identify A2744z8OD in this sample (Figure \ref{fig:A2744Cz8OD}). 
Moreover, there exist no other overdensities similar to A2744z8OD (Figure \ref{fig:distribution}). 
We confirm that A2744z8OD is a rare overdensity.
\item We calculate a galaxy surface overdensity $\delta$ of A2744z8OD with the homogenous $z\sim 8$ galaxy sample.
Defining the galaxy surface numbers of $\delta$ with a small circular area of $6''$ ($\simeq 30$ physical kpc) radius,
we find that $\delta$ of A2744z8OD is very large $\delta=132^{+66}_{-51}$. A2744z8OD is 
the most compact and dense galaxy overdensity at $z\gtrsim 2$ reported to date, 
and clearly different from those identified in the previous studies (Figure \ref{fig:chiang2013}).
\item We estimate the total stellar mass of the eight A2744z8OD galaxies with the mass-to-luminosity
		relation (Equation \ref{eq:stellar_mass}) to be $4 \times 10^9 M_\odot$
for \citet{2003PASP..115..763C} IMF.
This total stellar mass is as small as the one of today's Large Magellanic Cloud.
\item We search for a counterpart of A2744z8OD in the catalogs of Henriques2014a galaxy formation models,
after we confirm that the models reproduce the abundance and the clustering of the observed 
$z\sim 8$ dropout galaxies (Figures \ref{fig:surface_density} and \ref{fig:delta_distribution}).
In the models, we have found one overdensity, very similar to A2744z8OD,
that has eight model dropout galaxies at $z\sim8$ with a similar magnitude distribution to the one of observations (Figure \ref{fig:UV_hist}).
This model overdensity is referred to as Modelz8OD. 
Eight out of seven galaxies in Modelz8OD form a filament 
within $\Delta z \sim 0.03$ ($\sim 10^4$ km s$^{-1}$) 
that is elongated in the line of sight (Figure \ref{fig:display}). In addition to the
intrinsically high density in the three-dimensional space,
this filamentary distribution makes $\delta$ of Modelz8OD very high.
\item Investigating the evolution of Modelz8OD in the models, 
we have found that Modelz8OD is a progenitor of a today's cluster with a halo mass of $10^{14} M_\odot$ (Figure \ref{fig:mass_evolution}).
Moreover, four out of these seven Modelz8OD galaxies merge into
the brightest cluster galaxy (BCG) in the today's cluster.
If Modelz8OD is a similar system to A2744z8OD, the models suggest that
A2744z8OD would be a part of forming core of the today's cluster. 
A2744z8OD may be a laboratory for the early cluster core formation.
Because Modelz8OD is a rare ($\sim 2\sigma$) massive system at least up to 
$z\sim 12$ (Figure \ref{fig:stellar_evolution}), such a compact dense galaxy overdensity would 
trace the galaxy formation in the early universe.
\item Compact dense high-$z$ galaxy overdensities like A2744z8OD are unexplored in the past studies 
(Figure \ref{fig:chiang2013}).
Because such compact dense galaxy overdensities would be clue for understanding
early cluster core formation as well as the early galaxy formation,
one needs systematic studies of compact dense galaxy overdensities
that will complement popular studies of protoclusters defined by overdensities in a large-radius circle.
Future wide-field observation programs such as Subaru/Hyper Suprime-Cam surveys
will open up the studies of compact dense galaxy overdensities.

\end{enumerate}

\acknowledgments
We are grateful to Yoshiaki Ono for kindly providing us a part of HUDFP1 and P2 images and assisting our analyses. 
We also thank Ryota Kawamata and Masamune Oguri for creating our mass model and high-redshift galaxy catalogs as described in
\citet{2015ApJ...799...12I}.
We thank Mariko Kubo, Masayuki Umemura, Akio Inoue, Nobunari Kashikawa, Kazuhiro Shimasaku, Shun Saito, Daichi Kashino,
Akira Konno, Alex Hagen, Florent Duval
for valuable comments and discussion.
This study is partly conducted in the spring school program at Institute for Cosmic Ray Research in 2015 March.
We acknowledge useful insights obtained through the discussion with undergraduate students: 
Ryosuke Tominaga, Atsushi Yamamura, Ayano Shogaki, Hiroya Nishikawa, Ryotaro Sato, and Makiya Akagi.
This work is based on observations made with the NASA/ESA {\it Hubble Space Telescope}, 
obtained at the Space Telescope Science Institute (STScI),
which is operated by the Association of Universities for Research in Astronomy, Inc., 
under NASA contract NAS 5-26555.
The {\it HST} image mosaics were produced by the Frontier Fields Science Data Products Team at STScI.
This work was supported by World Premier International Research Center Initiative
(WPI Initiative), MEXT, Japan, and KAKENHI (23244025) and (15H02064) Grant-in-Aid 
for Scientific Research (A) through Japan Society for the Promotion of Science (JSPS).
The work of M.I. and Y.H. are partly supported by an Advanced Leading Graduate Course for Photon Science grant.

{\it Facilities:} \facility{{\sl HST} (WFC3, ACS)}


\bibliographystyle{apj}
\bibliography{apj-jour,ishigaki2015b}

\begin{thebibliography}{}
\expandafter\ifx\csname natexlab\endcsname\relax\def\natexlab#1{#1}\fi

\bibitem[{{Atek} {et~al.}(2015){Atek}, {Richard}, {Kneib}, {Jauzac},
  {Schaerer}, {Clement}, {Limousin}, {Jullo}, {Natarajan}, {Egami}, \&
  {Ebeling}}]{2015ApJ...800...18A}
{Atek}, H., {Richard}, J., {Kneib}, J.-P., {et~al.} 2015, \apj, 800, 18

\bibitem[{{Bertin} \& {Arnouts}(1996)}]{1996A&AS..117..393B}
{Bertin}, E., \& {Arnouts}, S. 1996, \aaps, 117, 393

\bibitem[{{Bolzonella} {et~al.}(2000){Bolzonella}, {Miralles}, \&
  {Pell{\'o}}}]{2000A&A...363..476B}
{Bolzonella}, M., {Miralles}, J.-M., \& {Pell{\'o}}, R. 2000, \aap, 363, 476

\bibitem[{{Boylan-Kolchin} {et~al.}(2009){Boylan-Kolchin}, {Springel}, {White},
  {Jenkins}, \& {Lemson}}]{2009MNRAS.398.1150B}
{Boylan-Kolchin}, M., {Springel}, V., {White}, S.~D.~M., {Jenkins}, A., \&
  {Lemson}, G. 2009, \mnras, 398, 1150

\bibitem[{{Bruzual} \& {Charlot}(2003)}]{2003MNRAS.344.1000B}
{Bruzual}, G., \& {Charlot}, S. 2003, \mnras, 344, 1000

\bibitem[{{Calzetti} {et~al.}(2000){Calzetti}, {Armus}, {Bohlin}, {Kinney},
  {Koornneef}, \& {Storchi-Bergmann}}]{2000ApJ...533..682C}
{Calzetti}, D., {Armus}, L., {Bohlin}, R.~C., {et~al.} 2000, \apj, 533, 682

\bibitem[{{Chabrier}(2003)}]{2003PASP..115..763C}
{Chabrier}, G. 2003, \pasp, 115, 763

\bibitem[{{Chiang} {et~al.}(2013){Chiang}, {Overzier}, \&
  {Gebhardt}}]{2013ApJ...779..127C}
{Chiang}, Y.-K., {Overzier}, R., \& {Gebhardt}, K. 2013, \apj, 779, 127

\bibitem[{{Coe} {et~al.}(2015){Coe}, {Bradley}, \&
  {Zitrin}}]{2015ApJ...800...84C}
{Coe}, D., {Bradley}, L., \& {Zitrin}, A. 2015, \apj, 800, 84

\bibitem[{{Curtis-Lake} {et~al.}(2016){Curtis-Lake}, {McLure}, {Dunlop},
  {Rogers}, {Targett}, {Dekel}, {Ellis}, {Faber}, {Ferguson}, {Grogin},
  {Kocevski}, {Koekemoer}, {Lai}, {M{\'a}rmol-Queralt{\'o}}, \&
  {Robertson}}]{2016MNRAS.457..440C}
{Curtis-Lake}, E., {McLure}, R.~J., {Dunlop}, J.~S., {et~al.} 2016, \mnras,
  457, 440

\bibitem[{{Driver} {et~al.}(2015){Driver}, {Davies}, {Meyer}, {Power},
  {Robotham}, {Baldry}, {Liske}, \& {Norberg}}]{2015arXiv150700676D}
{Driver}, S.~P., {Davies}, L.~J., {Meyer}, M., {et~al.} 2015, ArXiv e-prints,
  arXiv:1507.00676

\bibitem[{{Ellis} {et~al.}(2013){Ellis}, {McLure}, {Dunlop}, {Robertson},
  {Ono}, {Schenker}, {Koekemoer}, {Bowler}, {Ouchi}, {Rogers}, {Curtis-Lake},
  {Schneider}, {Charlot}, {Stark}, {Furlanetto}, \&
  {Cirasuolo}}]{2013ApJ...763L...7E}
{Ellis}, R.~S., {McLure}, R.~J., {Dunlop}, J.~S., {et~al.} 2013, \apjl, 763, L7

\bibitem[{{Franck} \& {McGaugh}(2015)}]{2015arXiv151204956F}
{Franck}, J.~R., \& {McGaugh}, S.~S. 2015, ArXiv e-prints, arXiv:1512.04956

\bibitem[{{Harikane} {et~al.}(2015){Harikane}, {Ouchi}, {Ono}, {More}, {Saito},
  {Lin}, {Coupon}, {Shimasaku}, {Shibuya}, {Price}, {Lin}, {Hsieh}, {Ishigaki},
  {Komiyama}, {Silverman}, {Takata}, {Tamazawa}, \&
  {Toshikawa}}]{2015arXiv151107873H}
{Harikane}, Y., {Ouchi}, M., {Ono}, Y., {et~al.} 2015, ArXiv e-prints,
  arXiv:1511.07873

\bibitem[{{Henriques} {et~al.}(2015){Henriques}, {White}, {Thomas}, {Angulo},
  {Guo}, {Lemson}, {Springel}, \& {Overzier}}]{2015MNRAS.451.2663H}
{Henriques}, B.~M.~B., {White}, S.~D.~M., {Thomas}, P.~A., {et~al.} 2015,
  \mnras, 451, 2663

\bibitem[{{Ishigaki} {et~al.}(2015){Ishigaki}, {Kawamata}, {Ouchi}, {Oguri},
  {Shimasaku}, \& {Ono}}]{2015ApJ...799...12I}
{Ishigaki}, M., {Kawamata}, R., {Ouchi}, M., {et~al.} 2015, \apj, 799, 12

\bibitem[{{Kawamata} {et~al.}(2015{\natexlab{a}}){Kawamata}, {Ishigaki},
  {Shimasaku}, {Oguri}, \& {Ouchi}}]{2015ApJ...804..103K}
{Kawamata}, R., {Ishigaki}, M., {Shimasaku}, K., {Oguri}, M., \& {Ouchi}, M.
  2015{\natexlab{a}}, \apj, 804, 103

\bibitem[{{Kawamata} {et~al.}(2015{\natexlab{b}}){Kawamata}, {Oguri},
  {Ishigaki}, {Shimasaku}, \& {Ouchi}}]{2015arXiv151006400K}
{Kawamata}, R., {Oguri}, M., {Ishigaki}, M., {Shimasaku}, K., \& {Ouchi}, M.
  2015{\natexlab{b}}, ArXiv e-prints, arXiv:1510.06400

\bibitem[{{Labb{\'e}} {et~al.}(2013){Labb{\'e}}, {Oesch}, {Bouwens},
  {Illingworth}, {Magee}, {Gonz{\'a}lez}, {Carollo}, {Franx}, {Trenti}, {van
  Dokkum}, \& {Stiavelli}}]{2013ApJ...777L..19L}
{Labb{\'e}}, I., {Oesch}, P.~A., {Bouwens}, R.~J., {et~al.} 2013, \apjl, 777,
  L19

\bibitem[{{Laporte} {et~al.}(2014){Laporte}, {Streblyanska}, {Clement},
  {P{\'e}rez-Fournon}, {Schaerer}, {Atek}, {Boone}, {Kneib}, {Egami},
  {Mart{\'{\i}}nez-Navajas}, {Marques-Chaves}, {Pell{\'o}}, \&
  {Richard}}]{2014A&A...562L...8L}
{Laporte}, N., {Streblyanska}, A., {Clement}, B., {et~al.} 2014, \aap, 562, L8

\bibitem[{{McLeod} {et~al.}(2015){McLeod}, {McLure}, {Dunlop}, {Robertson},
  {Ellis}, \& {Targett}}]{2015MNRAS.450.3032M}
{McLeod}, D.~J., {McLure}, R.~J., {Dunlop}, J.~S., {et~al.} 2015, \mnras, 450,
  3032

\bibitem[{{McLure} {et~al.}(2013){McLure}, {Dunlop}, {Bowler}, {Curtis-Lake},
  {Schenker}, {Ellis}, {Robertson}, {Koekemoer}, {Rogers}, {Ono}, {Ouchi},
  {Charlot}, {Wild}, {Stark}, {Furlanetto}, {Cirasuolo}, \&
  {Targett}}]{2013MNRAS.432.2696M}
{McLure}, R.~J., {Dunlop}, J.~S., {Bowler}, R.~A.~A., {et~al.} 2013, \mnras,
  432, 2696

\bibitem[{{Merlin} {et~al.}(2015){Merlin}, {Fontana}, {Ferguson}, {Dunlop},
  {Elbaz}, {Bourne}, {Bruce}, {Buitrago}, {Castellano}, {Schreiber}, {Grazian},
  {McLure}, {Okumura}, {Shu}, {Wang}, {Amor{\'{\i}}n}, {Boutsia}, {Cappelluti},
  {Comastri}, {Derriere}, {Faber}, \& {Santini}}]{2015A&A...582A..15M}
{Merlin}, E., {Fontana}, A., {Ferguson}, H.~C., {et~al.} 2015, \aap, 582, A15

\bibitem[{{Oesch} {et~al.}(2014){Oesch}, {Bouwens}, {Illingworth}, {Franx},
  {Ammons}, {van Dokkum}, {Trenti}, \& {Labbe}}]{2014arXiv1409.1228O}
{Oesch}, P.~A., {Bouwens}, R.~J., {Illingworth}, G.~D., {et~al.} 2014, ArXiv
  e-prints, arXiv:1409.1228

\bibitem[{{Oguri}(2010)}]{2010PASJ...62.1017O}
{Oguri}, M. 2010, \pasj, 62, 1017

\bibitem[{{Ono} {et~al.}(2013){Ono}, {Ouchi}, {Curtis-Lake}, {Schenker},
  {Ellis}, {McLure}, {Dunlop}, {Robertson}, {Koekemoer}, {Bowler}, {Rogers},
  {Schneider}, {Charlot}, {Stark}, {Shimasaku}, {Furlanetto}, \&
  {Cirasuolo}}]{2013ApJ...777..155O}
{Ono}, Y., {Ouchi}, M., {Curtis-Lake}, E., {et~al.} 2013, \apj, 777, 155

\bibitem[{{Salpeter}(1955)}]{1955ApJ...121..161S}
{Salpeter}, E.~E. 1955, \apj, 121, 161

\bibitem[{{Schenker} {et~al.}(2013){Schenker}, {Robertson}, {Ellis}, {Ono},
  {McLure}, {Dunlop}, {Koekemoer}, {Bowler}, {Ouchi}, {Curtis-Lake}, {Rogers},
  {Schneider}, {Charlot}, {Stark}, {Furlanetto}, \&
  {Cirasuolo}}]{2013ApJ...768..196S}
{Schenker}, M.~A., {Robertson}, B.~E., {Ellis}, R.~S., {et~al.} 2013, \apj,
  768, 196

\bibitem[{{Shibuya} {et~al.}(2015){Shibuya}, {Ouchi}, \&
  {Harikane}}]{2015ApJS..219...15S}
{Shibuya}, T., {Ouchi}, M., \& {Harikane}, Y. 2015, \apjs, 219, 15

\bibitem[{{Song} {et~al.}(2015){Song}, {Finkelstein}, {Ashby}, {Grazian}, {Lu},
  {Papovich}, {Salmon}, {Somerville}, {Dickinson}, {Duncan}, {Faber}, {Fazio},
  {Ferguson}, {Fontana}, {Guo}, {Hathi}, {Lee}, {Merlin}, \&
  {Willner}}]{2015arXiv150705636S}
{Song}, M., {Finkelstein}, S.~L., {Ashby}, M.~L.~N., {et~al.} 2015, ArXiv
  e-prints, arXiv:1507.05636

\bibitem[{{Springel} {et~al.}(2005){Springel}, {White}, {Jenkins}, {Frenk},
  {Yoshida}, {Gao}, {Navarro}, {Thacker}, {Croton}, {Helly}, {Peacock}, {Cole},
  {Thomas}, {Couchman}, {Evrard}, {Colberg}, \& {Pearce}}]{2005Natur.435..629S}
{Springel}, V., {White}, S.~D.~M., {Jenkins}, A., {et~al.} 2005, \nat, 435, 629

\bibitem[{{Toshikawa} {et~al.}(2014){Toshikawa}, {Kashikawa}, {Overzier},
  {Shibuya}, {Ishikawa}, {Ota}, {Shimasaku}, {Tanaka}, {Hayashi}, {Niino}, \&
  {Onoue}}]{2014ApJ...792...15T}
{Toshikawa}, J., {Kashikawa}, N., {Overzier}, R., {et~al.} 2014, \apj, 792, 15

\bibitem[{{Trenti} {et~al.}(2012){Trenti}, {Bradley}, {Stiavelli}, {Shull},
  {Oesch}, {Bouwens}, {Mu{\~n}oz}, {Romano-Diaz}, {Treu}, {Shlosman}, \&
  {Carollo}}]{2012ApJ...746...55T}
{Trenti}, M., {Bradley}, L.~D., {Stiavelli}, M., {et~al.} 2012, \apj, 746, 55

\bibitem[{{Venemans} {et~al.}(2007){Venemans}, {R{\"o}ttgering}, {Miley}, {van
  Breugel}, {de Breuck}, {Kurk}, {Pentericci}, {Stanford}, {Overzier}, {Croft},
  \& {Ford}}]{2007A&A...461..823V}
{Venemans}, B.~P., {R{\"o}ttgering}, H.~J.~A., {Miley}, G.~K., {et~al.} 2007,
  \aap, 461, 823

\bibitem[{{Zheng} {et~al.}(2014){Zheng}, {Shu}, {Moustakas}, {Zitrin}, {Ford},
  {Huang}, {Broadhurst}, {Molino}, {Diego}, {Infante}, {Bauer}, {Kelson}, \&
  {Smit}}]{2014ApJ...795...93Z}
{Zheng}, W., {Shu}, X., {Moustakas}, J., {et~al.} 2014, \apj, 795, 93

\end{thebibliography}

\end{document}